\documentclass[9pt,twocolumn,twoside]{osajnl}

\journal{jocn} 
\usepackage{comment}
\usepackage{graphicx} 
\usepackage{xcolor}
\usepackage{tikz}
\usetikzlibrary{arrows.meta,positioning,shapes.geometric,calc}
\usepackage{pgfplots}
\usepackage{pgfplotstable}
\usepackage{comment}
\usepackage{lettrine}
\pgfplotsset{compat=1.18}
\usepgfplotslibrary{groupplots}
\usetikzlibrary{patterns}

\setboolean{shortarticle}{false}

\title{When Less is More: Emerging Network-Wide Use Cases for Hollow-Core Fibers in  Optical Networks [Invited]}

\author{Mëmëdhe Ibrahimi\textsuperscript{*}}
\author{Giovanni Simone Sticca}
\author{Angelo Ferrara}
\author{\authorcr Francesco Musumeci}
\author{Massimo Tornatore}

\affil{Politecnico di Milano, Italy;  \textsuperscript{*}corresponding author: \href{mailto:memedhe.ibrahimi@polimi.it}{memedhe.ibrahimi@polimi.it}}

\begin{abstract}
Hollow-Core Fiber (HCF) is emerging as a disruptive technology for next-generation optical networks, offering ${\sim}30\%$ lower propagation latency than standard Single-Mode Fiber (SMF), record-low attenuation (0.05--0.1~dB/km), and nonlinearities reduced by three to four orders of magnitude. While the physical-layer benefits of HCF have been extensively demonstrated, recent industrial commitments by hyperscalers (e.g., Microsoft, Amazon) and the rapid maturation of nested antiresonant nodeless fibers (NANF) call for a systematic assessment of the impact of HCF at the network level. Building on a series of prior contributions, we overview four emerging network-wide use cases for the \emph{selective deployment of HCF in hybrid SMF/HCF networks}: (i) latency-constrained Edge Data Center (edgeDC) consolidation; (ii) energy-efficient high-power amplification; (iii) capacity scaling via single-fiber bidirectional (BiDi) transmission; and (iv) coexistence of classical and quantum (QKD) channels. Consolidated results across realistic metro, national, and continental topologies convey a unifying message: in each use case, a selective HCF upgrade captures a significant share of the achievable benefit compared to a network overhaul. The required HCF budget is use-case-specific, i.e., $\sim$6--24\% of the total network fiber length for edgeDC consolidation (up to 50\% fewer edgeDCs), $\sim$40\% of the network links for QKD/classical coexistence (up to 38\% fewer quantum modules), and $\sim$40--60\% of the network links for energy-efficient capacity scaling via BiDi-HCF (up to 28\% lower network power per Tbps and 70\% more served traffic, at half the HCF strand count of a unidirectional HCF upgrade), whereas for energy-efficient high-power amplification the gain keeps growing with the HCF budget (up to 60\% lower network power per Tbps at the national scale) and what saturates is the benefit of increasing the EDFA output power. Finally, we discuss limitations of current network-wide HCF studies, open operational challenges, and emerging opportunities for HCF.
\end{abstract}

\setboolean{displaycopyright}{false} 

\begin{document}

\maketitle

\section{Introduction}
\lettrine{H}{ollow-Core} Fiber (HCF) represents a paradigm shift in optical communications, overcoming the fundamental limits of solid-core silica fibers. While the concept of guiding light in an air core was demonstrated in the late 1990s~\cite{knight1998photonic}, the recent advances in Nested Antiresonant Nodeless Fibers (NANF) have allowed HCF to surpass the long-standing attenuation records of standard Single-Mode Fibers (SMF), reaching values between 0.05 and 0.1~dB/km~\cite{petrovich2025nature,chen2024ofc,ding2025ecoc}. Beyond the \emph{hero transmission loss}, HCF offers additional distinctive characteristics: (i) a propagation latency that is 30\% lower than SMF (because light propagates in air rather than silica), and (ii) nonlinearity that is reduced by three to four orders of magnitude with respect to SMF~\cite{poggiolini2022jlt}.

These properties have moved HCF from a laboratory curiosity to a strategic technology in industrial roadmaps. In 2022, Microsoft acquired Lumenisity, a pioneer in HCF manufacturing, and in 2025 announced plans to scale production through partnerships with Corning and Heraeus~\cite{Deployment2,microsoft_forster_2025}. In parallel, Amazon Web Services has outlined a roadmap for the integration of HCF in key Availability Zones to support latency-sensitive workloads, in particular AI-driven traffic~\cite{callaghan_reinvent24}, and China Mobile has demonstrated HCF deployment for Data Center Interconnects (DCI)~\cite{xiong2024ecoc}. The wave of industrial interest is 
matched by increased academic activity, with HCF-related contributions at top optical-communication venues such as OFC and ECOC, and an emerging set of network-wide studies~\cite{poggiolini2022jlt,iqbal2021,iqbal2023ofc,pedro2025ecoc,correia2025ondm,correia2025ofc,zami2025ofc,zami2025apc}.

While fiber design, fabrication processes, and physical-layer transmission experiments have been demonstrated
~\cite{petrovich2025nature}, the focus is now shifting towards the complete network ecosystem, raising the question: 
\emph{how} and \emph{where} should HCF be deployed in optical networks? This is a non-trivial question because: 1) HCF is unlikely to fully replace SMF in the medium term, due to the limited production capacity, the immaturity of the supply chain, and the still-high (and uncertain) cost per km. Therefore, hybrid SMF/HCF networks, where HCF is used \emph{selectively} on a subset of links, are the most plausible deployment scenario; 2) The selection of \emph{which} links to upgrade may depend on the targeted use case: minimizing the number of edge data centers (edgeDCs) under latency constraints may lead to upgrade patterns that are different from those that minimize energy per bit or maximize served traffic; 
and 3) Different use cases may benefit from different HCF types (e.g., low-attenuation HCF for capacity scaling, low-noise/wideband HCF for QKD coexistence, low-backscatter HCF for bidirectional transmission), suggesting that, unlike SMF (where ITU-T G.652 dominates), HCF deployment may be intrinsically segment- and use-case-specific.

\noindent This paper is intended as an overview: it consolidates, in a self-contained form, network-level results published separately in~\cite{sticca2024ofc,sticca2024ecoc,sticca2025ofc,sticca2026tnsm,ibrahimi2025ondm,sticca2026jocn,sticca2026ondm,ibrahimi2026ecoc}, it builds on, and follows the logic of, the invited talk presented at OFC~2026, and it extracts cross-cutting insights from the consolidated results. With respect to our prior work, 
it adds: (i) the lexicographic tier-aware objective and the edgeDC-disjoint variant of $\S$3.\ref{sec:edgeDC}; 
(ii) the network-wide analysis of BiDi-HCF on the EU49 topology ($\S$3.\ref{sec:bidirectional}); 
and (iii) the discussion of the scope of validity of the selective-deployment observation, and of the limitations, open challenges, and opportunities of HCF ($\S$\ref{sec:usecases} and$\S$\ref{sec:challenges}). 
The four use cases span metro, regional, national, and continental segments, and cover latency, capacity, energy efficiency, and security-oriented requirements. Specifically, the contributions of this paper are summarized as follows:

\begin{enumerate}
    \item We provide an overview of the state of the art of HCF, covering fiber design and production, practical deployments by hyperscalers and telecom operators, and network-level studies. We provide a summary table with the main dimensions along which HCF has progressed and the use cases for which HCF has been shown to be useful ($\S$\ref{sec:sota}).
    \item We review and consolidate four use cases for the \emph{selective} deployment of HCF in next-generation optical networks ($\S$\ref{sec:usecases}), namely:
    \begin{enumerate}
        \item \emph{Edge Data Center consolidation} in latency-constrained metro/regional networks~\cite{sticca2024ofc,sticca2026tnsm};
        \item \emph{Energy-efficient high-power amplification} in continental and national topologies, leveraging the ultra-low nonlinearity of HCF and emerging energy-efficient ZR/ZR+ pluggable transponders~\cite{sticca2024ecoc,sticca2026jocn};
        \item \emph{Capacity scaling} via hybrid SMF/HCF networks and via the emerging opportunity of single-fiber bidirectional HCF (BiDi-HCF) transmission~\cite{sticca2025ofc,ibrahimi2026ecoc};
        \item \emph{Coexistence of classical and quantum (QKD) channels} in metro-area networks
        (HCF removes the spontaneous Raman scattering penalty constraining SMF~\cite{sticca2026ondm}).
    \end{enumerate}
    For each use case, we present the problem statement, methodology, and representative numerical results. 

    
    \item We discuss the \emph{limitations} of current network-wide HCF studies (including ours), the \emph{open challenges} that the technology faces (manufacturing scale-up, standardization, field deployment, operations, aging), and the \emph{opportunities} that HCF unlocks beyond the four use cases analyzed in the paper, including access and 6G RAN, subsea systems, and unconventional bands ($\S$\ref{sec:challenges}).
\end{enumerate}
Taken together, the four use cases show that the advantages of HCF from having \emph{less} i) attenuation, ii) nonlinearity, iii) propagation latency, iv) Rayleigh backscattering, and v) Raman scattering, compared to SMF, result in \emph{more} i) reach within a latency budget, ii) launch power and hence more GSNR, and iii) more capacity per fiber strand. 
Moreover, under the assumptions considered, a limited and well-placed amount of HCF captures a significant share of the benefits of a network-wide overhaul. The latter is an observation on the analyzed scenarios rather than a general rule (see  $\S$~\ref{sec:usecases} and in $\S$4.~\ref{sec:limitations}).

\section{State of the Art of Hollow-Core Fibers}
\label{sec:sota}

\subsection{Advances in HCF Design and Production}~\label{sec:sota_fiber}
HCF research dates back to the late 1990s, when Knight \emph{et al.}~\cite{knight1998photonic} demonstrated photonic-bandgap guidance of light in a hollow channel. Since then, HCF has evolved through photonic-bandgap and antiresonant fibers to NANF, Double-NANF (DNANF), and support-tube (ST-HCF) variants~\cite{poletti2025ofc}. The key figure of merit driving this evolution, attenuation, has fallen by an order of magnitude in less than a decade, from $\sim$1.3~dB/km (2018) to $0.040\pm0.003$~dB/km (2026)~\cite{chen2024ofc,petrovich2025nature,ding2025ecoc,gao2025ecoc,li2026ofc}. Table~\ref{tab:sota} details this progression, together with the parallel improvements in Inter-Modal Interference (IMI), now below the $-60$~dB/km penalty threshold~\cite{poggiolini2022jlt,li2026ofc} (and further mitigated at the system level by digital subcarrier multiplexing~\cite{ospina2026ofc}), and in SMF--HCF splicing, down to 0.15--0.05~dB with automated, field-deployable procedures~\cite{suslov2021,feng2026ofc,zhang2025ecoc_ar,slavik2026ofc}.
Two aspects remain critical for network-scale deployment, and we return to them in $\S$~\ref{sec:challenges}: \emph{i)} fabrication \emph{scalability}, where record single-draw lengths of 83~km~\cite{gao2025ecoc} and closed-loop draw control~\cite{budd2026ofc} are narrowing (but not yet closing) the gap between \emph{hero-loss} and volume-production losses (0.147~dB/km batch-averaged over 733~km~\cite{ding2025ecoc}); and \emph{ii)} gas-line absorption, i.e., narrow CO$_2$-induced dips in the C/L band that are intrinsic to the air core~\cite{petrovich2025nature}. A further review of HCF design and fabrication is beyond the scope of this paper, and we refer the interested reader to~\cite{poletti2025ofc,petrovich2025nature,ge2026jlt}.

\begin{table*}[t]
\caption{Summary of the state of the art on HCF: progress and use cases. Entries marked $\star$ are contributions of the authors that are consolidated and extended in this paper.}
\label{tab:sota}
\centering
\footnotesize
\begin{tabular}{p{2.4cm} p{4.0cm} p{6.0cm} p{3.6cm}}
\toprule
\textbf{Dimension} & \textbf{Recent progress} & \textbf{Representative result} & \textbf{Key references} \\
\midrule
Attenuation & DNANF, support-tube and gap-tube-assisted HCF, broadband NANF & 0.174~dB/km (2022), $\leq$0.11~dB/km (2024), 0.05~dB/km (2025), 0.040~dB/km (2026) & \cite{chen2024ofc,petrovich2025nature,ding2025ecoc,gao2025ecoc,li2026ofc} \\
Inter-Modal Interference & Improved core/cladding geometry, mode stripping, gap-tube designs & Reduced from $-35$ to $-63.6$/$-73.2$~dB/km, below the $-60$~dB/km penalty threshold & \cite{poggiolini2022jlt,li2026ofc} \\
SMF--HCF splicing & Low-loss, anti-reflection-coated, field-deployable interconnection & From $\sim$0.15~dB to 0.05~dB splice loss; back-reflection below $-33$~dB & \cite{suslov2021,feng2026ofc,zhang2025ecoc_ar,slavik2026ofc} \\
Fabrication scalability & Interstitial/gap-tube geometries, closed-loop draw control & Record 83~km single-draw span; but 0.147~dB/km batch-averaged over 733~km & \cite{gao2025ecoc,budd2026ofc,ding2025ecoc} \\
Unrepeated / ultra-long reach & High-power amplification, low loss, low IMI & $>$300~km unamplified; 266~km spans in a 6660~km transoceanic link & \cite{ali2025ofc,boddeda2026ofc} \\
Field deployment & DCI and AZ-to-AZ links; operator field trials & Microsoft (production), Amazon (roadmap), Verizon (field study), China Mobile ($\leq$42.7~km cables, $<$0.1~dB/km, 114.9~Tb/s) & \cite{microsoft_forster_2025,fang2026ofc,ge2026jlt,ge2026ofc} \\
Manufacturing scale-up & Outsourced manufacturing, strategic partnerships & Microsoft/Corning/Heraeus, $>$15{,}000~km planned in 24~months & \cite{Deployment2} \\
\midrule
\multicolumn{4}{l}{\textbf{Network-level use cases}} \\
\midrule
Latency-constrained DC consolidation & Joint edgeDC \& HCF placement, MILP/MINLP, GA & Up to 50\% fewer edgeDCs; a ${\sim}6$--24\% HCF budget already suffices (saturates) & \cite{sticca2024ofc,sticca2026tnsm,ibrahimi2026ofc}$^{\star}$ \\
High-power amplification \& energy efficiency & Joint EDFA/TXP power, ZR/ZR+ pluggables & Up to 33\% (EU) and 60\% (IT) lower W/Tbps; a moderate $P_{\max}$ (${\sim}29$~dBm) already captures most gains at national scale & \cite{sticca2024ecoc,sticca2026jocn}$^{\star}$,~\cite{zami2026ofc} \\
Capacity scaling & Hybrid SMF/HCF (C, C+L); span placement under CAPEX budget & Up to $2.1\times$ traffic at $\sim$25\% lower cost; hybrid spans match full-HCF reach & \cite{sticca2025ofc}$^{\star}$,~\cite{zami2025ofc,pedro2026ofc,braga2026ofc} \\
Bidirectional HCF for capacity scaling & Selective BiDi-HCF placement, low Rayleigh backscatter & BiDi-HCF matches Uni-HCF at half the HCF strands; up to ${\sim}28\%$ lower W/Tbps and ${\sim}70\%$ more served traffic, with a ${\sim}40$--60\% HCF budget & \cite{ibrahimi2026ecoc}$^{\star}$ \\
QKD/classical coexistence & C-band quantum + classical co-propagation & Up to 38\% fewer quantum modules at a 40\% HCF budget (saturates) & \cite{sticca2026ondm}$^{\star}$ \\
Long-haul throughput & C+L, ultra-wideband, Raman-assisted & 1.5$\times$--5$\times$ throughput vs.\ SMF & \cite{poggiolini2022jlt,zami2025ofc,zami2025apc} \\
\bottomrule
\end{tabular}
\end{table*}

\subsection{Practical Deployments and Industrial Roadmaps}
\label{sec:sota_deploy}
The maturation of HCF has triggered a rapid industrial uptake. Microsoft Azure has reported the first commercial deployments of HCF in its production network for latency-critical DCI routes~\cite{microsoft_forster_2025}, and announced plans for outsourced, large-scale HCF manufacturing through agreements with Corning and Heraeus~\cite{Deployment2}. Amazon AWS has presented a roadmap to integrate HCF in Availability Zones supporting latency-sensitive AI workloads~\cite{callaghan_reinvent24}. On the operator side, the most extensive field activity has been reported by China Mobile, in collaboration with fiber manufacturers Linfiber and YOFC. 
Building on earlier field deployments~\cite{xiong2024ecoc,gao2025ecoc_cable}, they have deployed AR-HCF of 10--42.7~km in an urban network with an average link loss of 0.113~dB/km, demonstrating real-time terabit-scale S$+$C$+$L transmission~\cite{ge2026jlt} and real-time single-fiber bidirectional transmission of 128~Tb/s~\cite{ge2025ofc_fd}.
In parallel, a first field study of a deployed AR-HCF cable in a Verizon DCI has recently been reported~\cite{fang2026ofc}. 

This wave of industrial activity indicates that HCF has reached the maturity required to carry production traffic in selected and well-controlled segments, in particular short-reach DCI and intra-region links between availability zones. It represents a first step rather than a general validation: wide-scale field experience, multi-year reliability data, and standardized specifications are not yet available ($\S$4.\ref{sec:open_challenges}). It nonetheless suggests that the research focus should broaden from \emph{whether} HCF works in the field to \emph{where} and \emph{how} HCF should be deployed at the network level.

\begin{figure*}[t]
    \centering
    \includegraphics[width=0.85\textwidth]{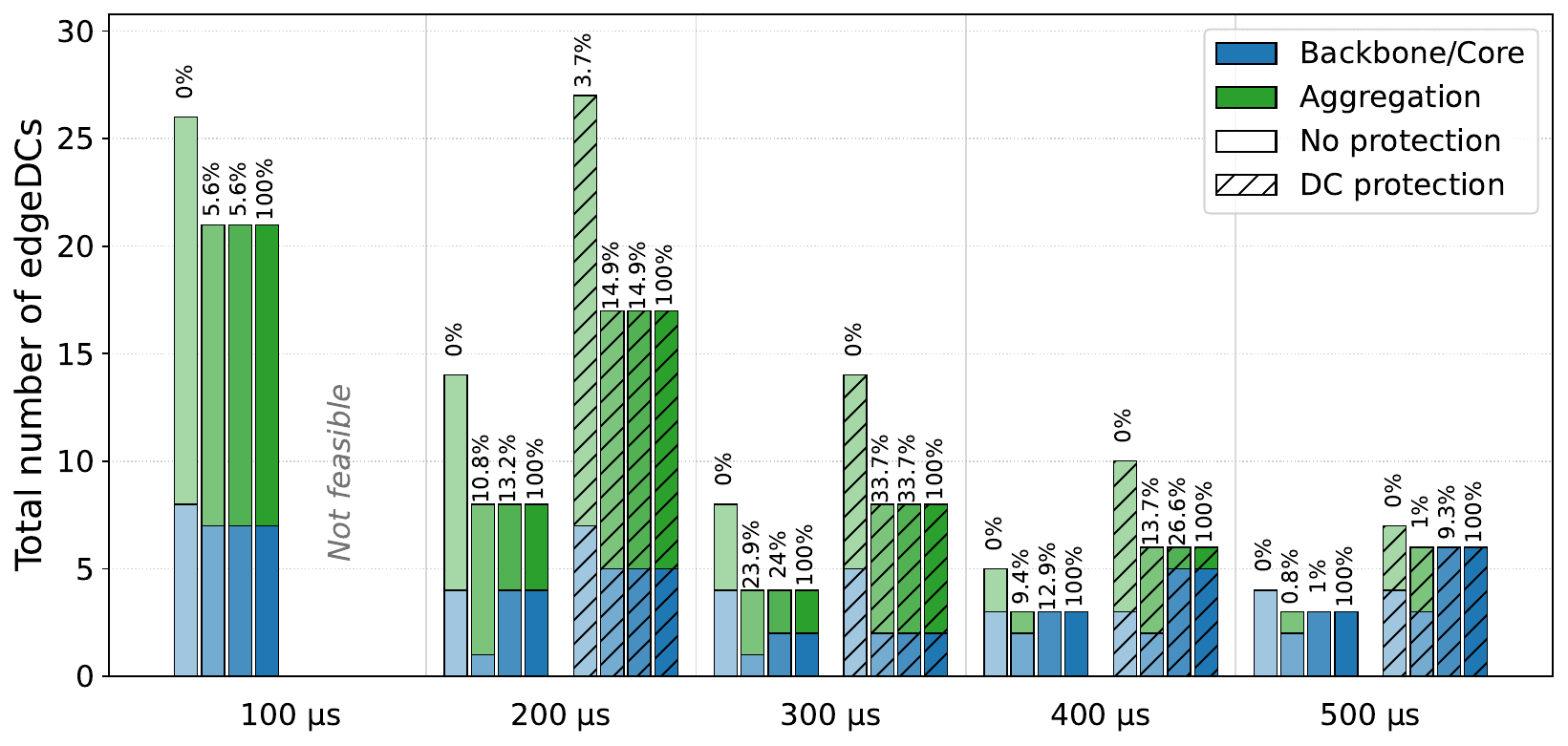}
    \caption{Total number of deployed edgeDCs versus client-to-edgeDC latency threshold (100--500~$\mu s$), for HCF budgets between 0\% (SMF-only) and 100\% (HCF-only), in the \textit{Metro-52} topology, for the \emph{No protection} and \emph{edgeDC-disjoint assignment} scenarios.} 
    \vspace{-10pt}
    \label{fig:edgeDC-metro52}
\end{figure*}

\subsection{System- and Network-Level Studies on HCF} \label{sec:sota_network}
A first wave of network-level HCF studies focused on long-haul throughput and reach. Poggiolini and Poletti~\cite{poggiolini2022jlt} established a comprehensive analytical framework for HCF transmission, showing throughput gains of $1.5\times$--$5\times$ over benchmark SMF C+L Raman-amplified systems for link distances from 100 to 4000~km, and identifying $-60$~dB/km as the IMI threshold for negligible system penalty. 
Most recently, the same authors made the case for a bidirectional DNANF trans-Atlantic submarine cable approaching 1~Pb/s per direction with 200~km spans~\cite{poggiolini2025ecoc}. 

A second strand of research analyzed HCF in operator-grade backbone scenarios. Iqbal \emph{et al.}~\cite{iqbal2021,iqbal2023ofc} demonstrated 400ZR DWDM transmission over field-deployed HCF and analyzed the feasibility of HCF in UK metro/regional networks. 
Pedro and Correia~\cite{pedro2025ecoc,correia2025ondm,correia2025ofc} explored selective HCF placement in regional/long-haul networks, showing that HCF can reduce the number of inline amplifiers and meet performance targets even under realistic IMI and amplifier constraints, and, most recently, formulated HCF-span upgrades under an explicit CAPEX budget~\cite{pedro2026ofc}.
Zami \emph{et al.}~\cite{zami2025ofc,zami2025apc} benchmarked HCF against optimized SMF in transparent WDM backbones, highlighting that the magnitude of the gains is strongly sensitive to baseline assumptions (SMF loss, span length, and amplification scheme), and later showed that inline EDFAs delivering more than 30--32~dBm are not energy-optimal~\cite{zami2026ofc}. 
At the link level, Braga \emph{et al.}~\cite{braga2026ofc} found that a hybrid HCF-SMF span can match a full-HCF system of equal reach while replacing up to half of the HCF in metro systems. In addition, Saber and Jiang~\cite{saber2026ecoc_prot} investigated protection switching in hybrid SMF/HCF networks. 


Table~\ref{tab:sota} summarizes the state of the art, along both the dimensions of physical-layer and industrial progress (attenuation, IMI, splicing, fabrication scalability, reach, field deployment, and manufacturing scale-up) and the network-level use cases for which HCF has been shown to be beneficial. 
Rather than re-detailing the individual studies in Table~\ref{tab:sota}, this paper 
examines four use cases, for each of which we assess the value of a \emph{selective} HCF deployment consistent with current industrial realities.
The community is independently converging on this selective-deployment regime: the most recent studies constrain the amount of HCF deployed~\cite{pedro2026ofc,braga2026ofc} and identify moderate operating points~\cite{zami2026ofc}, corroborating the observation that motivates this paper, namely that a limited, well-placed HCF upgrade captures a significant share of its network-level benefit, while also confirming that the magnitude of the benefit, and the budget at which it saturates, depend on the underlying assumptions~\cite{zami2025apc}.

\section{Network-Wide Use Cases for HCF}
\label{sec:usecases}
We present four \emph{high-value use cases} for the selective deployment of HCF. For each use case, we follow the same structure: \emph{problem definition}, \emph{methodology and evaluation settings}, and \emph{representative numerical results}. 
The objective is not to repeat in detail the modeling already published in~\cite{sticca2024ofc,sticca2024ecoc,sticca2025ofc,sticca2026tnsm,ibrahimi2025ondm,sticca2026jocn,sticca2026ondm}, but to provide a self-contained synthesis and to highlight cross-cutting insights.

The \emph{HCF budget} denotes the fraction of the network upgraded to HCF and is measured using two metrics. In the edgeDC use case ($\S$3.\ref{sec:edgeDC}), it is the fraction of the total fiber length upgraded to HCF (i.e., the MILP selects links of any length in a metro/aggregation topology with strongly heterogeneous link lengths). In the other use cases ($\S$3.\ref{sec:hpa}--$\S$3.\ref{sec:qkd}), it is the fraction of network links upgraded to HCF (links are selected by a topology-driven ranking or by the genetic algorithm in $\S$3.\ref{sec:qkd}).

Unless otherwise stated, HCF is modeled with a 0.11~dB/km attenuation and a $-60$~dB/km IMI~\cite{poggiolini2022jlt}. The attenuation value is deliberately not a hero value: it matches the average link loss measured on field-deployed AR-HCF cables (0.113~dB/km~\cite{ge2026jlt}) and lies between the batch-averaged production loss (0.147~dB/km~\cite{ding2025ecoc}) and the current \emph{hero loss} records (0.040~dB/km~\cite{li2026ofc}). The IMI value is in line with related studies~\cite{poggiolini2022jlt} and is already improved by the latest fibers ($-63.6$ to $-73.2$~dB/km~\cite{li2026ofc}). The lumped losses (0.5~dB per connector, 0.7~dB per circulator, and 1.0~dB per MUX/DEMUX) are conservative with respect to the SMF--HCF splice losses of Table~\ref{tab:sota}.

\emph{Infrastructure versus provisioning decisions.} The methodologies of $\S$3.B and $\S$3.C separate infrastructure decisions from provisioning decisions, in view of their different lifetimes (decades for the fiber plant and the amplifier sites, months to years for lightpaths and transponders). The fiber-type assignment, and in $\S$3.B the inline-EDFA placement, are computed once and offline from topology-level information (edge betweenness centrality) and are therefore independent of any specific set of demands, whereas routing, modulation and spectrum assignment, and transponder selection are evaluated a posteriori on the fixed infrastructure, over many random traffic realizations and with a single transponder generation, so that the reported figures characterize the infrastructure rather than a single demand set. This sequential approach trades some optimality for robustness: a joint optimization for a known traffic matrix could yield a better infrastructure (see~\cite{pedro2026ofc} for traffic-aware span placement), at the risk of tailoring a decades-long investment to demands that change within years. Moreover, since several transponder generations will succeed one another over the lifetime of the fiber plant, the budgets at which the benefit flattens and the energy-optimal $P_{\max}$ should be expected to shift over time, and a multi-period planning framework accounting for traffic growth and transponder evolution is a natural direction for future work.


\subsection{HCF for Edge Data Center Consolidation} \label{sec:edgeDC}
\subsubsection{Problem definition and motivation}
Latency-sensitive services (5G/6G, cloud gaming, extended reality, distributed AI training and inference, autonomous driving) require that compute resources be placed within a strict propagation delay from the user. In SMF-based metro/regional networks, this propagation budget translates into a maximum geographic ``coverage radius'' for each edgeDC. Reducing the propagation latency by 30\%, as enabled by HCF, expands this coverage radius and reduces the number of edgeDCs required to serve a given population. Because each edgeDC entails substantial CAPEX and OPEX (real estate, power, cooling, IT hardware, operations), reducing the number of edgeDCs has a direct economic and environmental impact.

The latency threshold is the \emph{propagation-delay budget} between a client and its serving edgeDC. 
Other fixed contributions to the end-to-end latency (transponder DSP and FEC processing, node processing, queuing, and processing at the edgeDC) are independent of the fiber type, and as they take a larger share of the end-to-end budget, the results are reported for a range of propagation budgets (100--500~$\mu$s). 

The \emph{edgeDC network design with HCF} problem~\cite{sticca2024ofc,sticca2026tnsm} can be formally stated as: \emph{given} a network topology with candidate links, a set of clients with associated client-to-edgeDC latency thresholds, client-to-edgeDC and inter-edgeDC traffic demands, a set of candidate edgeDC locations, and an HCF budget (i.e., the maximum fiber length that can be upgraded to HCF); \emph{decide} (i) the placement of edgeDCs and (ii) the placement of HCF links; \emph{subject to} client-to-edgeDC latency constraints; 
\emph{with the objective} of minimizing the number of deployed edgeDCs.

\subsubsection{Methodology}
We adopt the Mixed Integer Linear Programming (MILP) formulation proposed in~\cite{sticca2024ofc}, which jointly optimizes the placement of HCF links and edgeDCs subject to an HCF upgrade budget 
and a per-client latency threshold. 
With respect to~\cite{sticca2024ofc}, we introduce a lexicographic refinement of the objective to account for node importance. The primary objective still aims to minimize the total number of deployed edgeDCs. Among all solutions achieving this minimum, we then progressively minimize the number of edgeDCs assigned to lower-priority node types, i.e., the MILP is biased toward placing edgeDCs on Core/Backbone nodes rather than Aggregation nodes whenever multiple placements achieve the same minimum edgeDC count.
Additionally, we evaluate an \emph{edgeDC-disjoint} variant of the MILP, in which each client is assigned to two edgeDCs, both within the latency threshold, so that service can be sustained upon the failure of either edgeDC. No additional constraint is imposed on the disjointness of the underlying network paths (path protection is orthogonal and can be layered on top).


\begin{figure*}[t]
    \centering
    \includegraphics[width=0.9\columnwidth]{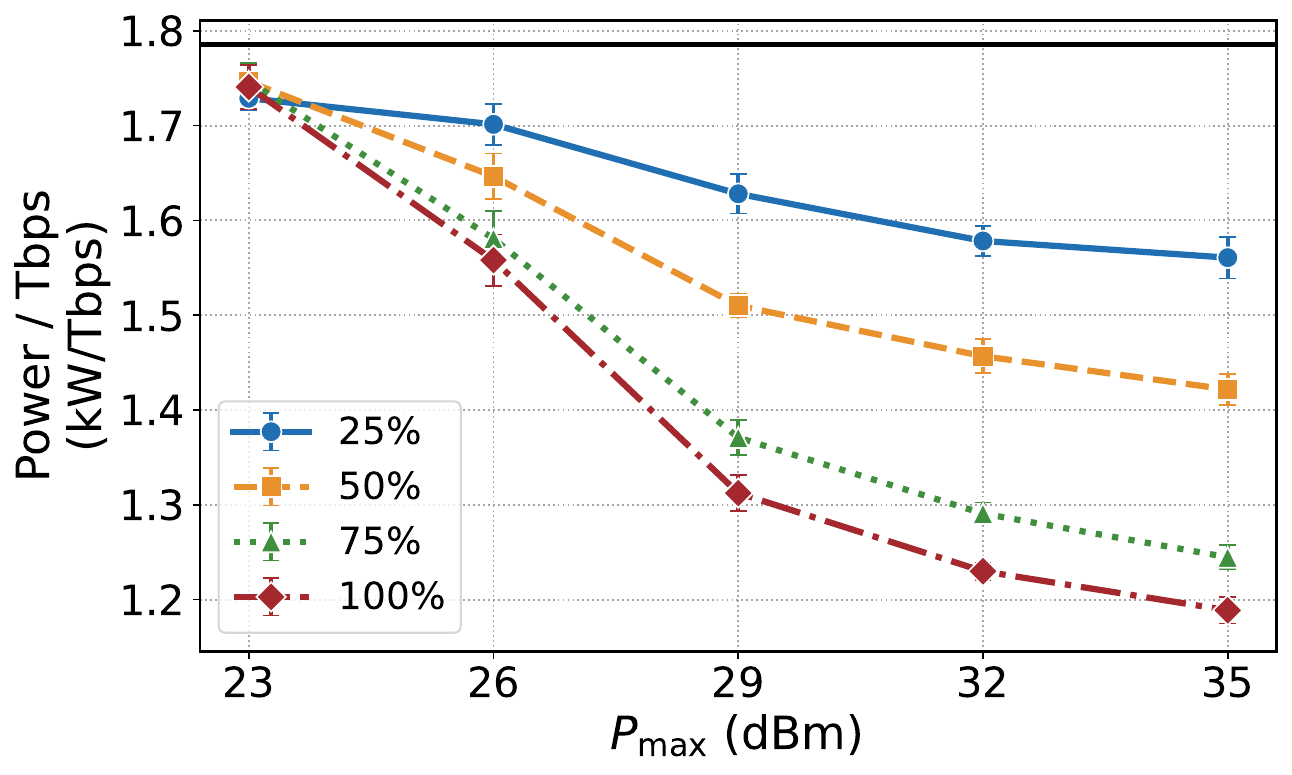}
    \includegraphics[width=0.9\columnwidth]{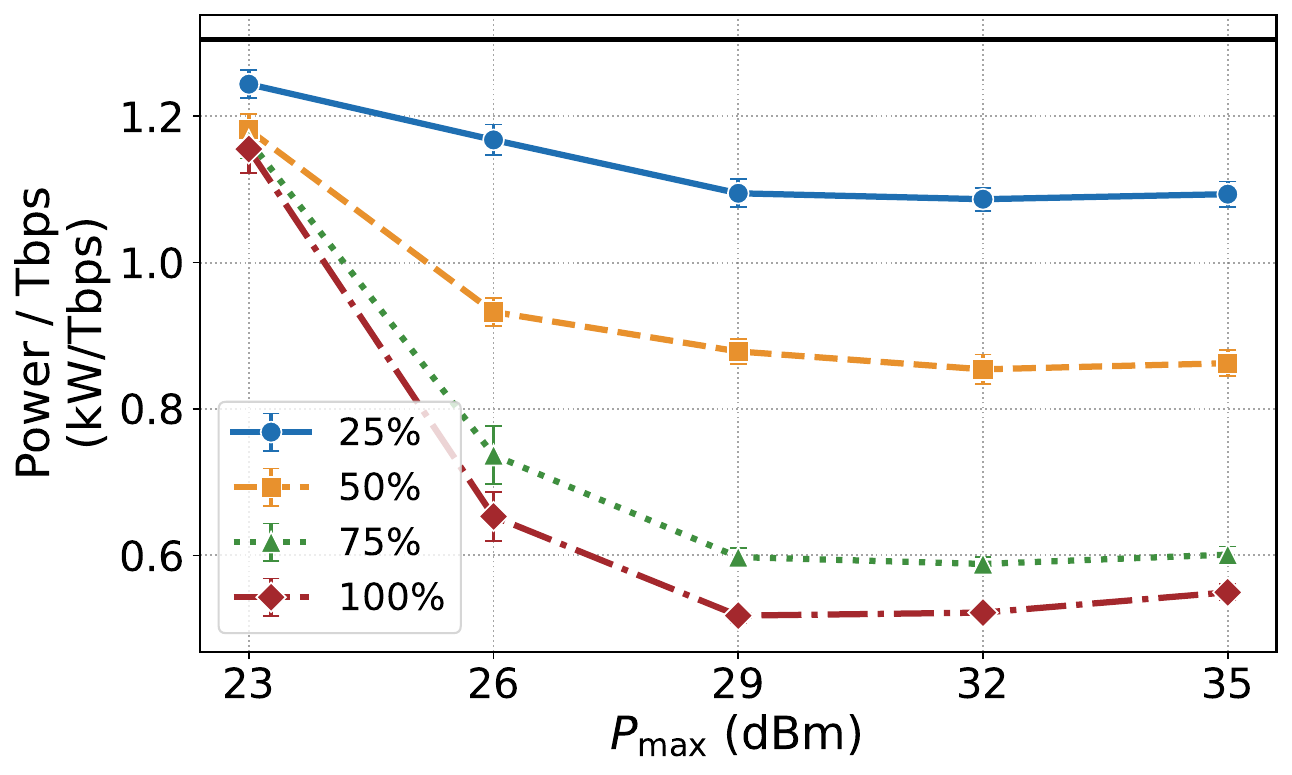}
    \vspace{-10pt}
    \caption{Total network power consumption per Tbps (kW/Tbps) versus maximum EDFA output power $P_{\max}$, for HCF budgets of 25--100\% of upgraded links. (Left) Continental EU topology. (Right) National IT topology.}
    \label{fig:highpower-EU}
\end{figure*}

\subsubsection{Representative results}
We evaluate the edgeDC consolidation approach on a 52-node metro/aggregation (\textit{Metro-52}~\cite{sticca2024ofc}) topology, sweeping the HCF budget in $[0\%,100\%]$ and the client-to-edgeDC latency threshold in $[100,500]~\mu s$. The topology includes eight Core/Backbone nodes and 44 Aggregation nodes (see Fig.~6 in~\cite{9082151}).

Fig.~\ref{fig:edgeDC-metro52} illustrates the numerical results for the unprotected and edgeDC-disjoint scenarios. 
Let us first consider the scenario without protection. 
Even a small HCF budget delivers substantial benefits: 
the minimum HCF budget needed to reach the same edgeDC \emph{count} achievable with a 100\% HCF network stays below $\sim$24\% across the entire latency range (peaking at 300~$\mu s$, and as low as a few percent at the tightest and loosest thresholds). 
However, matching not only the edgeDC count but also the same tier-biased node \emph{distribution} as the 100\% HCF budget solution (i.e., edgeDCs preferentially placed on Core/Backbone nodes rather than Aggregation nodes) requires a slightly larger budget, e.g., at 200~$\mu s$, 10.8\% of HCF budget suffices to match the edgeDC count, and 13.2\% are needed to also reproduce the same node distribution. Within this range, the number of deployed edgeDCs drops by up to \emph{50\%}, for an average reduction of \emph{38\%} across the latency range. Beyond these thresholds, additional HCF budget yields no further reduction, i.e., the consolidation gains are effectively saturated.

Considering the edgeDC-disjoint assignment, we observe that it affects \emph{feasibility}, i.e., the solution must ensure that each client reaches two edgeDCs while respecting the latency constraint, not just edgeDC count. The 100~$\mu s$ threshold becomes infeasible even at a 100\% HCF budget, since no pair of independently reachable edgeDCs exists for every client under such a tight constraint, and at 200~$\mu s$ the edgeDC-disjoint assignment is in turn infeasible at a 0\% HCF budget. Where feasible (200--500~$\mu s$), it roughly doubles the number of required edgeDCs with respect to the unprotected case (e.g., from 8 to 17 at 200~$\mu s$). However, introducing HCF in the edgeDC-disjoint network still yields substantial savings, and the gains saturate in the same way as in the unprotected case, albeit at a higher budget: the minimum HCF budget needed to reach the fully-upgraded edgeDC count grows to $\sim$15--34\%, depending on the latency threshold, versus 6--24\% in the unprotected case.

\emph{Relation to other studies.} 
Other network-level studies on the propagation latency use case 
(e.g., Iqbal \emph{et al.}~\cite{iqbal2023ofc}), 
do not consider the joint placement of compute resources and HCF. 
The practical value of HCF for latency has also been demonstrated by hyperscalers~\cite{microsoft_forster_2025,callaghan_reinvent24}) to extend the latency-bounded reach of their regions (by $\times$2.25 compared to an SMF-only scenario).

\subsection{HCF for Energy-Efficient High-Power Amplification}
\label{sec:hpa}

\subsubsection{Problem definition and motivation}
SMF-based optical networks are constrained in their launch power by Kerr nonlinearity: beyond a given launch power 
the GSNR is dominated by nonlinear interference (NLI) rather than amplified spontaneous emission (ASE) noise, so that a further increase of the amplifier output power is detrimental.

In HCF, where nonlinearity is reduced by 3--4 orders of magnitude, 
EDFA output powers up to 38~dBm have been considered in the literature~\cite{poggiolini2022jlt}, opening a new design degree of freedom: \emph{higher launch power means higher GSNR}, which translates either into longer reach (fewer/longer spans, fewer inline EDFAs) or into higher spectral efficiency. 
The challenge becomes \emph{when and to what extent} the EDFA output power should be increased: a higher per-amplifier electrical draw must be balanced against the savings enabled by SNR improvements.

We formulate the \emph{energy-efficient high-power network design in hybrid HCF/SMF networks}~\cite{sticca2026jocn} problem as: \emph{given} a topology with a fiber type assignment, a set of TXPs (including ZR/ZR+ pluggables), a maximum EDFA output power $P_{\max}$, and a set of traffic demands, \emph{decide} the routing, modulation, spectrum assignment, TXP type, and inline EDFA placement, \emph{subject to} SNR feasibility, spectrum, and amplifier-gain constraints, \emph{with the objective} of jointly maximizing served traffic and minimizing total network power (EDFAs and TXPs).

\subsubsection{Methodology}
The methodology consists of three steps~\cite{sticca2026jocn}: (step 1) preprocessing, in which a target HCF budget $\rho \in [0,1]$ is mapped to the fiber type assignment by ranking links via a $k$-shortest-path traversal-frequency centrality over all node pairs (i.e., edge betweenness centrality) and upgrading the top-$\rho |L|$ links; (step 2) the ROMA (eneRgy-efficient Optical aMplifier plAcement) heuristic, which iteratively removes inline EDFAs on HCF links subject to feasibility (gain limits, $P_{\max}$, SNR) and objective improvement in terms of power consumption (EDFA+TXP power); 
(step 3) traffic evaluation via incremental RMSA up to 1\% blocking. 
A detailed description of the methodology can be found in ~\cite{sticca2026jocn}. 
In this paper (and in~\cite{sticca2026jocn}) we extend our previous ECOC~2024 model~\cite{sticca2024ecoc} along two key dimensions: (i) we include transponder power consumption (which often dominates over the EDFA contribution), modeled from vendor datasheets~\cite{infinera_chm,zhang2024jocn}, and include 400G~ZR+ pluggable modules; (ii) we consider \emph{hybrid} SMF/HCF deployments with HCF budgets of 25\%, 50\%, 75\%, and 100\%, in addition to SMF-only case. The considered transponder set comprises long-haul transponders operating at 150, 200, 250, and 300~Gb/s (8QAM, 16QAM, 32QAM, and 64QAM, with GSNR thresholds from 15.8 to 24.0~dB) and 400G ZR+ pluggables (GSNR threshold of 24.3~dB). The complete portfolio, with symbol rates, spectral occupations, and power consumptions, is listed in Table~2 of~\cite{sticca2026jocn}.


 
\subsubsection{Representative results}
We evaluate the high-power amplification approach on two reference topologies of increasing geographical scale and average link length: 
a 19-node continental European network (EU19) with an average link length of $\sim$655~km, and a 21-node national Italian network (IT21) with an average link length of $\sim$181~km.
We vary the maximum EDFA output power $P_{\max}$ over $\{23, 26, 29, 32, 35\}$~dBm on HCF links, while SMF links retain $P_{\max}=23$~dBm. 
Fig.~\ref{fig:highpower-EU} reports the total network power consumption normalized to the served traffic (in kW/Tbps) as a function of $P_{\max}$ and of the HCF budget (25\%, 50\%, 75\%, and 100\% of network links upgraded to HCF), for the EU and the IT topology. The horizontal line indicates the \emph{SMF-only} baseline.
 
\emph{Continental EU topology}. The SMF-only baseline sits at $\sim$1.78~kW/Tbps. Increasing the HCF budget consistently reduces the power consumption per Tbps, and the benefit grows steadily with $P_{\max}$. At $P_{\max}=29$~dBm, moving from SMF-only to 25\% HCF already yields a $\sim$9\% reduction, 
while a full HCF upgrade reaches up to $26\%$ savings.  
At $P_{\max}=35$~dBm, the 50\% HCF configuration reaches $20\%$ savings, and the full HCF upgrade reaches $33\%$ savings. 
The kW/Tbps curves decrease across the $P_{\max}$ range, confirming that high-power operation is consistently beneficial in a long-haul continental setting. 
 
\emph{National IT topology}. The SMF-only baseline sits at $\sim$1.31~kW/Tbps. The relative gains are substantially larger than in the EU topology, thanks to the shorter link lengths, which make lower-power ZR/ZR+ pluggables feasible on a larger fraction of lightpaths. 
At $P_{\max}=29$~dBm, moving from SMF-only to 25\% HCF gives a $\sim$17\% reduction, while a full HCF upgrade brings the power consumption down by $60\%$. 

Interestingly, the IT curves saturate much earlier than the EU curves: beyond $P_{\max}=29$~dBm, further increases in EDFA output power do not yield additional energy-efficiency gains. This early saturation is a property of the considered transponder set 
rather than of the topology alone. On the shorter links of the IT topology, with HCF budgets of 50\% or more, the average GSNR of the lightpaths already exceeds at $P_{\max}=26$--29~dBm the 24.3~dB required by the highest-rate mode of the considered transponders, so that the additional GSNR made available by a higher $P_{\max}$ can no longer be exploited through transponder selection, while the removal of inline EDFAs brings limited savings on links of $\sim$181~km average length. On the EU topology, instead, the average GSNR crosses the thresholds of the higher-rate modes only between 26 and 35~dBm, so that every increase of $P_{\max}$ moves additional lightpaths to a higher-rate, more energy-efficient mode. 
Transponders supporting higher-order or probabilistically shaped modulation formats and higher symbol rates 
would continue to exploit the additional GSNR at higher $P_{\max}$ on the IT topology too, and the IT curves would then be expected to follow a trend closer to the EU ones, up to the point where the back-to-back SNR of the transceiver, rather than the link, limits the achievable gain~\cite{poggiolini2022jlt,klaus2022ofc}.


\emph{Relation to other studies.} Our finding that a moderate $P_{\max}$ ($\sim$26--29~dBm) is suitable at the national scale is consistent with the transparent-network study of Zami \emph{et al.}~\cite{zami2026ofc}, and with the reduction of inline amplifiers obtained by Correia and Pedro~\cite{correia2025ondm} through selective HCF upgrades. 
The continental-scale trend, where the benefit continues to grow up to $P_{\max}=35$~dBm, is in turn consistent with the reach-oriented use of high launch powers ~\cite{ali2025ofc,mardoyan2026ofc}. 

\begin{figure*}[t]
    \centering
    \includegraphics[width=\linewidth]{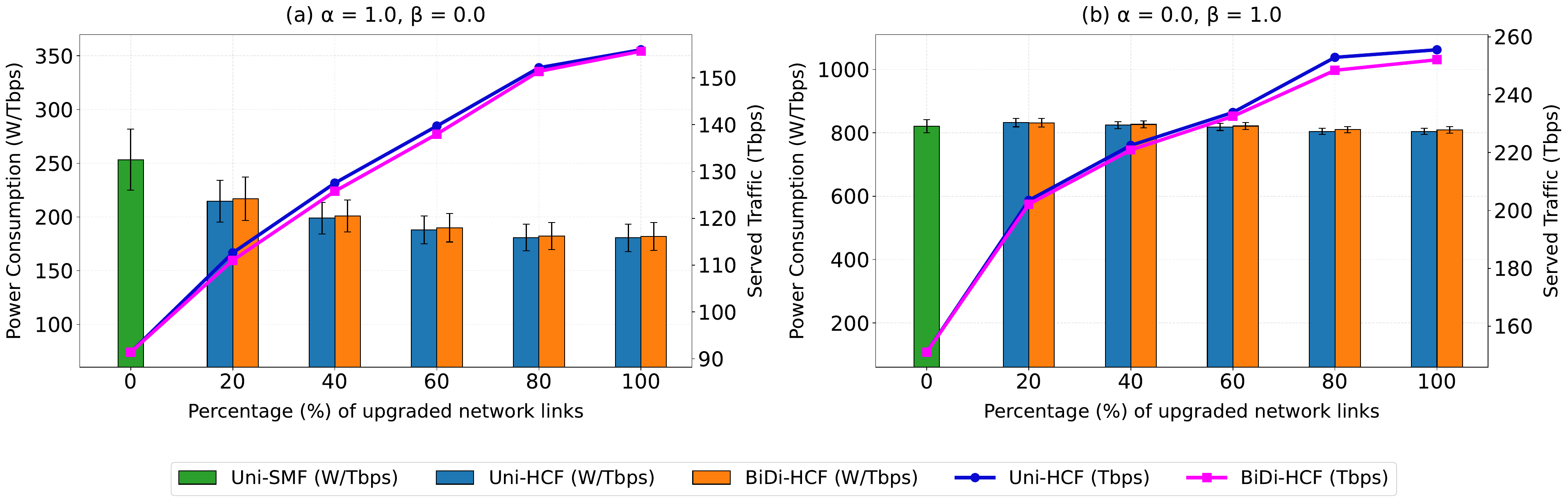}
    \caption{Power consumption per Tbps (bars) and total served traffic (lines) in the EU49 topology as a function of the percentage of upgraded network links, for (a) $\alpha=1$, $\beta=0$ (power-oriented) and (b) $\alpha=0$, $\beta=1$ (spectrum-oriented).
    }
    \label{fig:bidir-results}
\end{figure*}
\begin{figure}[t]
    \centering
    \includegraphics[width=0.99\linewidth]{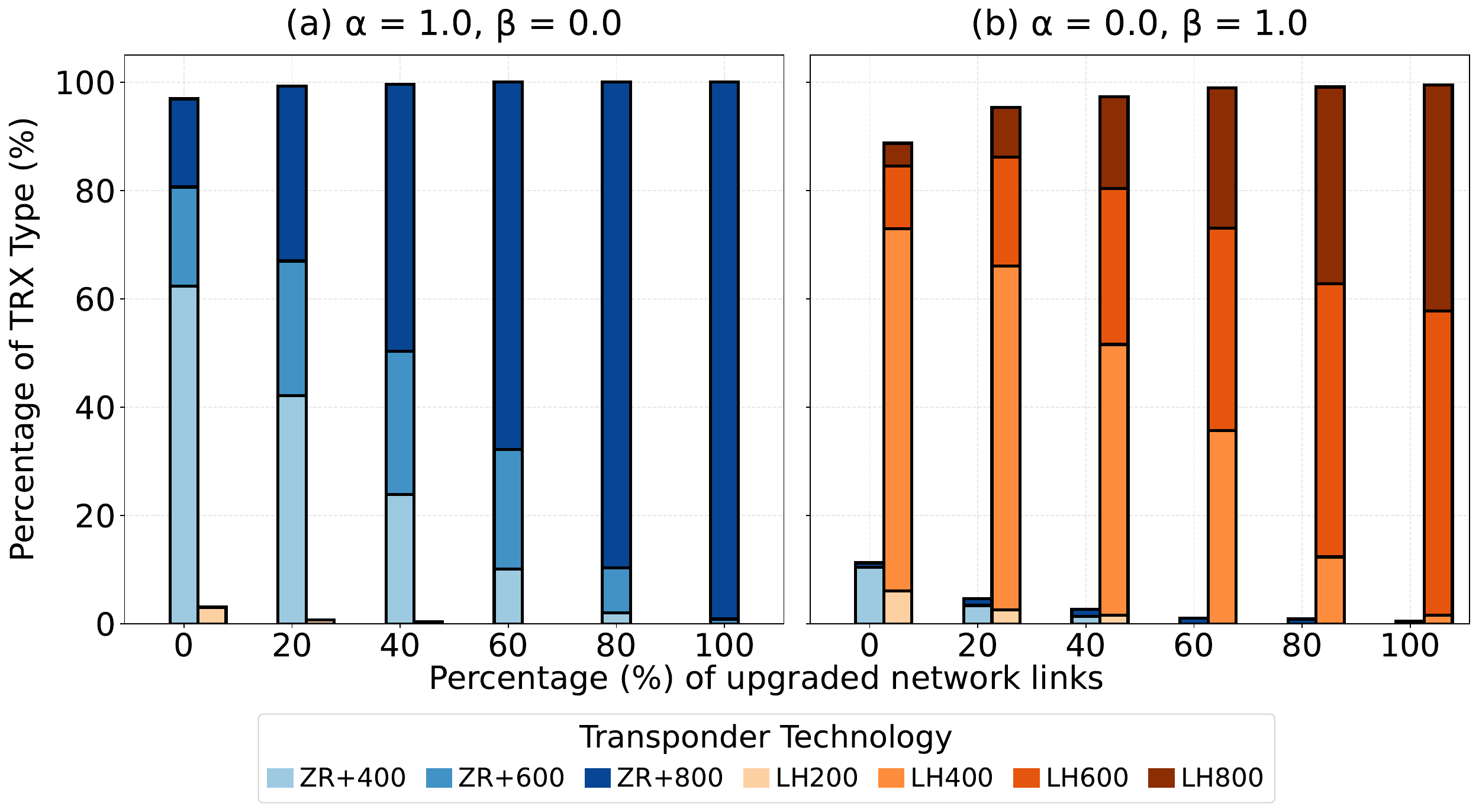}
    \caption{TXP distribution in the BiDi-HCF scenario, under (a) power-oriented and (b) spectrum-oriented TXP selection.}
    \vspace{-15pt}
    \label{fig:bidir-txp}
\end{figure}

\subsection{HCF for Energy-Efficient Capacity Scaling}
\label{sec:bidirectional}
 
\subsubsection{Problem definition and motivation}
Traffic in optical transport networks continues to grow at $\sim$30--40\% per year, driven by cloud, 5G/6G, and AI-related workloads. 
Several enablers are explored to scale capacity~\cite{sticca2025ofc}, namely, parallel SMF fibers, multi-band transmission, multi-core fibers, and HCF, each with distinct trade-offs: parallel fibers grow linearly in cost (more TXPs and WSSes), multi-band requires band-specific amplification with non-trivial impairments, and multi-core requires specialized amplifiers and connectors.
In~\cite{sticca2025ofc}, we investigated HCF as a long-term capacity-scaling technology in \emph{unidirectional} hybrid SMF/HCF networks, comparing selective HCF upgrades in C- and (C+L)-band against the ``natural'' capacity-scaling alternative of deploying $n$ parallel SMF fibers. 
 
Since then, the HCF technology has matured significantly: recent industrial announcements~\cite{Deployment2} and physical-layer demonstrations~\cite{hong2025ecoc,ding2025ecoc,poggiolini2025ecoc} have consolidated HCF as a credible medium for capacity scaling. This maturity has, in turn, shifted the research question from ``\emph{is HCF a competitive capacity-scaling option?}'' to a more nuanced ``\emph{how should HCF be deployed so as to simultaneously minimize CAPEX, minimize power consumption, and maximize throughput?}'' 
Regarding this, a particularly compelling opportunity is the exploitation of \emph{single-fiber bidirectional transmission} (BiDi-HCF). While SMF-based BiDi transmission is severely limited by Rayleigh backscattering (which causes coherent crosstalk between the two propagation directions), in HCF the Rayleigh backscattering coefficient is approximately 30~dB lower than in SMF~\cite{Slavik:22,Zhang:25}, enabling high-performance BiDi transmission over a single fiber. From a network-planning perspective, BiDi-HCF is particularly attractive because it \emph{simultaneously} addresses three cost/energy challenges that are unique to HCF: (i) it doubles the per-fiber capacity at the cost of additional in-line equipment, e.g., circulators, an extra CAPEX that is easily justified for an expensive fiber such as HCF; (ii) it halves the number of HCF strands needed at the network level, a non-trivial gain given that low-loss HCF designs typically require a larger cladding diameter than SMF~\cite{poletti2025ofc}; 
and (iii) it preserves the energy-efficiency gains of Uni-HCF, allowing the network to keep operating on low-power ZR/ZR+ pluggables.
 
Building on this observation, in our recent work~\cite{ibrahimi2026ecoc}, we formulate the \emph{Selective BiDi-HCF Deployment} problem as follows. \emph{Given} a network topology, a set of traffic demands, a set of available TXPs, and an HCF budget (i.e., the maximum number of network links that can be upgraded to HCF); \emph{decide} the fiber-type assignment (Uni-SMF, Uni-HCF, or BiDi-HCF), the RMSA of demands, and the TXP type; \emph{subject to} physical-layer feasibility (including bidirectional backscattering-induced crosstalk on BiDi-HCF links) and spectrum continuity/contiguity constraints; \emph{with the objective} of jointly minimizing network power consumption and maximizing served traffic.
 
\subsubsection{Methodology}
\begin{figure}[t]
\centering
    \includegraphics[width=0.99\linewidth]{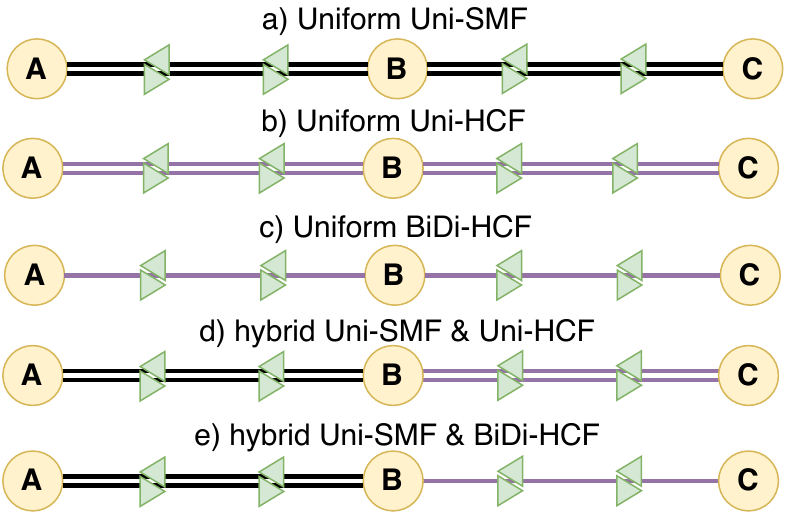}
    \caption{Network-wide deployment scenarios. 
    Hybrid Uni-SMF\,\&\,Uni-HCF and Hybrid Uni-SMF\,\&\,BiDi-HCF allow a fraction $\rho$ of links to be upgraded. 
    }
    \label{fig:bidir-scenarios}
\end{figure}

We consider three link-level deployments: \emph{(i) Uni-SMF}: a pair of unidirectional SMF fibers (the legacy baseline); \emph{(ii) Uni-HCF}: a pair of unidirectional HCF fibers; and \emph{(iii) BiDi-HCF}: a single HCF strand carrying BiDi transmission on the same spectrum. These link-level deployments give rise to five network-level scenarios (Fig.~\ref{fig:bidir-scenarios}): (1) \emph{Uniform Uni-SMF}, our legacy baseline; (2) \emph{Uniform Uni-HCF}; (3) \emph{Uniform BiDi-HCF}; (4) \emph{Hybrid Uni-SMF \& Uni-HCF}, where a fraction $\rho$ of links is upgraded to Uni-HCF; and (5) \emph{Hybrid Uni-SMF \& BiDi-HCF}, where a fraction $\rho$ of links is upgraded to BiDi-HCF. 
In the hybrid scenarios, links to be upgraded are selected via \emph{edge betweenness centrality}, computed as in $\S$3.\ref{sec:hpa}, i.e., top-$\rho$ links are chosen.
\begin{table}[t]
\centering
\caption{Transponder catalog corresponding characteristics, including Long-haul (LH) and ZR+ pluggable.} 
\label{tab:txp}
\footnotesize
\setlength{\tabcolsep}{3pt}
\resizebox{\columnwidth}{!}{%
\begin{tabular}{l c c c c c}
\toprule
\textbf{Type} & \textbf{Bit rate} & \textbf{Symbol rate} & \textbf{Slots} & \textbf{GSNR th.} & \textbf{Power} \\
 & \textbf{[Gb/s]} & \textbf{[GBd]} & \textbf{(12.5~GHz)} & \textbf{[dB]} & \textbf{[W]} \\
\midrule
LH200~\cite{infinera_chm,Oleg}  & 200 & 62.5 & 6  & 5.4  & 120 \\
LH400~\cite{infinera_chm,Oleg}  & 400 & 62.5 & 6  & 11.5 & 164 \\
LH600~\cite{infinera_chm,Oleg}  & 600 & 62.5 & 6  & 16.7 & 208 \\
LH800~\cite{infinera_chm,Oleg}  & 800 & 84   & 8  & 16.6 & 251 \\
ZR+400~\cite{yang_ecoc_25,chang_ecoc_25} & 400 & 131 & 12 & 7.0  & 23 \\
ZR+600~\cite{yang_ecoc_25,chang_ecoc_25} & 600 & 131 & 12 & 11.3 & 27 \\
ZR+800~\cite{yang_ecoc_25,chang_ecoc_25} & 800 & 131 & 12 & 13.1 & 31 \\
\bottomrule
\end{tabular}}
\vspace{-10pt}
\end{table}
For each scenario, we solve the RMSA problem with $k$-shortest-path routing ($k=3$) and first-fit flex-grid spectrum with 12.5~GHz slot spacing in the C-band. 
Among all feasible transmission modes on the candidate path, the TXP type is chosen, among the LH transponders and ZR+ pluggables of Table~\ref{tab:txp}, by minimizing a weighted objective ($\alpha + \beta=1$) that balances TXP power consumption ($\alpha=1$, $\beta=0$) and spectrum-slot occupation ($\alpha=0$, $\beta=1$). 
Physical-layer modeling adopts~\cite{semrau2019} for SMF ($0.22$~dB/km) and~\cite{poggiolini2022jlt} for HCF, with the BiDi-HCF crosstalk model of~\cite{poggiolini2025ecoc} (Rayleigh backscattering coefficient of $-100$~dB/m, validated in~\cite{Zhang:25}), per-span lumped losses of 0.5~dB per connector, 0.7~dB per circulator, and 1.0~dB per MUX/DEMUX. 


\subsubsection{Representative results}
We evaluate the five scenarios on the EU49 continental topology (49 nodes, 138 unidirectional links, $\sim$393~km average link length), generating traffic requests from $\{400, 800, 1200, 1600\}$~Gbps with probabilities $\{0.25, 0.40, 0.25, 0.10\}$, and incrementally loading the network until 1\% blocking is reached. 
We consider the LH transponders and ZR+ pluggables of Table~\ref{tab:txp}. 
Each configuration is averaged over 100 independent runs. Since the focus of this use case is on energy-efficient capacity scaling, we primarily report results under the power-oriented TXP-selection regime ($\alpha=1$, $\beta=0$), and briefly contrast them with the spectrum-oriented regime ($\alpha=0$, $\beta=1$) at the end of this section.

Fig.~\ref{fig:bidir-results}(a) reports the power consumption per Tbps (bars, left axis) and the total served traffic (lines, right axis) as a function of the HCF-upgrade percentage, under the power-oriented regime. The main observations are: (a) all HCF-integrated scenarios reduce W/Tbps and increase the served traffic. The Uniform Uni-SMF baseline sits at $\sim$253~W/Tbps and serves $\sim$91~Tbps. Progressively upgrading the network with HCF (either Uni-HCF or BiDi-HCF) reduces the power consumption per Tbps, i.e., a 20\% HCF upgrade reduces W/Tbps by $\sim$14\% and a full HCF upgrade by $\sim$28\% (down to $\sim$182~W/Tbps), while the served traffic grows by $\sim$70\% (up to $\sim$155~Tbps). 
Both effects are driven by the adoption of the highest-rate ZR+ mode on a growing share of lightpaths. 
(b) BiDi-HCF matches Uni-HCF, at half the HCF strand count. The hybrid BiDi-HCF and hybrid Uni-HCF curves track each other closely (within $\sim$2\%, in both metrics) across the full range of HCF budgets, which means that \emph{BiDi-HCF achieves essentially the same W/Tbps and served traffic as Uni-HCF while deploying half the number of HCF fiber strands}, effectively neutralizing the CAPEX and duct-space penalties associated with HCF; (c) \emph{A partial upgrade captures most of the gains for the considered transponder set.} Most of the W/Tbps reduction is achieved within the first 40--60\% of upgraded links: a 40\% BiDi-HCF budget captures roughly three quarters, and a 60\% budget about 90\%, of the reduction achievable with a full upgrade, whose floor is $\sim$182~W/Tbps. 
This floor, and the budget at which it is approached, are set by the transponder portfolio: the reduction is driven by the migration of lightpaths toward the highest-rate ZR+ mode (ZR+800, Fig.~\ref{fig:bidir-txp}(a)), and once the HCF budget allows nearly all lightpaths to be served at 800G ($\sim$90\% of them at an 80\% budget), additional HCF cannot lower the W/Tbps further. Transponders with a lower power per bit or a longer reach would lower the floor and shift the budget at which it is approached. The served traffic, instead, keeps growing up to an 80\% HCF budget (from $\sim$91~Tbps to $\sim$151~Tbps).

The underlying mechanism is illustrated in Fig.~\ref{fig:bidir-txp}, which reports the TXP distribution. Under the power-oriented regime (Fig.~\ref{fig:bidir-txp}(a)), ZR+ pluggables serve $\sim$97\% of the lightpaths already in the Uni-SMF network (LH200 transponders take the remaining $\sim$3\%), so that the HCF upgrade acts on the mode selected within the ZR+ class: the share of ZR+800 grows from $\sim$16\% in Uni-SMF to $\sim$49\% at a 40\% budget, $\sim$90\% at 80\%, and $\sim$99\% at full HCF, while ZR+400 drops from $\sim$62\% to $\sim$0\%. Since ZR+800 carries twice the rate of ZR+400 in the same 12 slots for $\sim$35\% more power (Table~\ref{tab:txp}), this migration drives both the W/Tbps reduction and the growth of the served traffic.

For completeness, Fig.~\ref{fig:bidir-results}(b) shows the spectrum-oriented regime ($\alpha=0$, $\beta=1$), which instead prioritizes maximum throughput at the expense of power efficiency. Here, the served traffic grows with the HCF budget (from $\sim$151~Tbps for Uni-SMF to $\sim$255~Tbps at 100\% HCF), and BiDi-HCF again matches Uni-HCF at half the strand count. However, the network now relies predominantly on high-rate LH TXPs (Fig.~\ref{fig:bidir-txp}(b), which serve $\sim$90\% of the lightpaths in Uni-SMF and $\sim$99\% at full HCF, shifting from LH400 ($\sim$72\% in Uni-SMF) toward LH600 and LH800 ($\sim$56\% and $\sim$41\% at full HCF)), so that, compared to the power-oriented regime, the power consumption per Tbps is 3--4.5 times higher and essentially independent of the HCF budget. The spectrum-oriented regime thus serves $\sim$65\% more traffic in absolute terms, owing to the lower spectral occupation of LH TXPs, but at a much higher energy cost per bit. 
This confirms that for a network operator jointly targeting energy efficiency and capacity scaling, the power-oriented regime combined with a selective BiDi-HCF deployment on 40--60\% of links is the most attractive configuration.

\begin{figure}[t]
\centering
    \includegraphics[width=1.0\columnwidth]{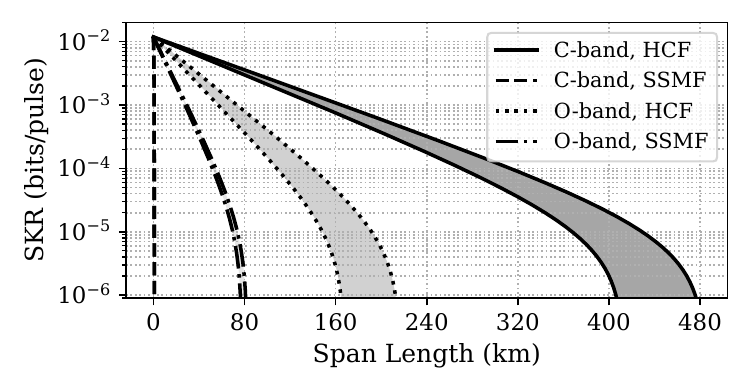}
    \caption{Secret Key Rate (SKR) vs. Span Length (km), for HCF and SMF in case of C- and O-band.}
    \vspace{-10pt}
    \label{fig:example-qkd}
\end{figure}

\begin{figure}[t]
\centering
    \includegraphics[width=1.0\columnwidth]{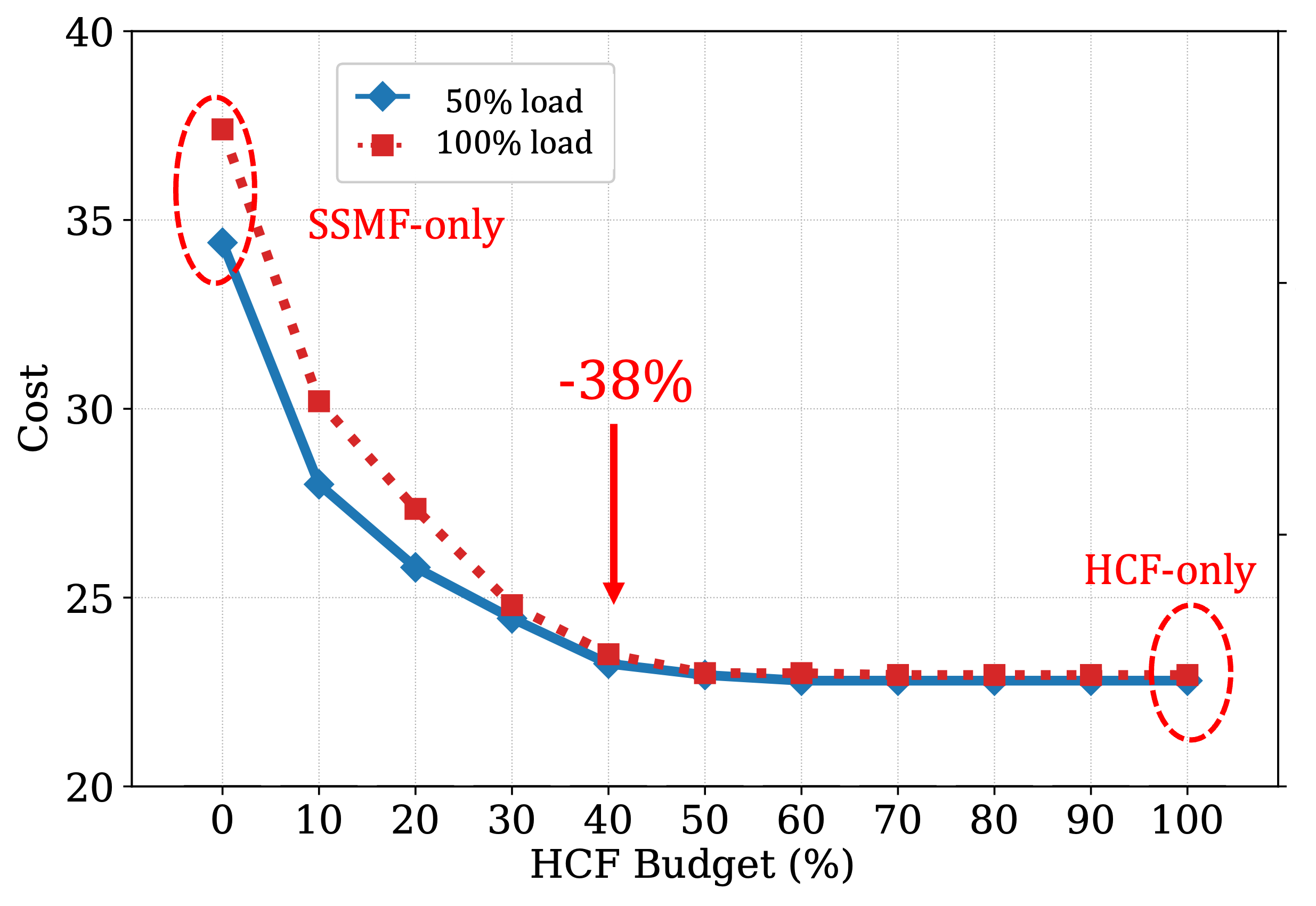}
    \caption{Average cost (number of deployed quantum modules) as a function of the HCF budget, under 50\% (half-load) and 100\% (full-load) of classical channels.}
    \vspace{-10pt}
    \label{fig:result-qkd}
\end{figure}
 

\emph{Relation to other studies.} The observation that a moderate HCF budget already captures a large share of the attainable gain is consistent with the CAPEX-constrained placement study of Pedro \emph{et al.}~\cite{pedro2026ofc}. BiDi transmission over HCF with commercial 400ZR/800ZR pluggables has been demonstrated in~\cite{hong2026ofc_bidi,dupas2026ofc}, and S+C+L bidirectional transmission beyond 550~Tb/s in~\cite{li2026ofc_bidi}, supporting the physical-layer feasibility of the BiDi-HCF scenario. These system-level demonstrations pave the way for network-level studies that quantify the benefits of BiDi-HCF across the optical network ecosystem.

\subsection{HCF for QKD and Classical Communication Coexistence} ~\label{sec:qkd}
\subsubsection{Problem definition and motivation}
Quantum Key Distribution (QKD) is a promising technology to provide information-theoretic security, with steady progress in deployment. A practical challenge is the \emph{coexistence} of QKD channels with classical (high-power) WDM channels on the same fiber: classical channels generate spontaneous Raman scattering (SpRS) noise that, at conventional-band wavelengths, can easily overwhelm the single-photon-level quantum channel.

In SMF, the standard mitigation is to transmit the quantum channel in the O-band (where SpRS from C-band classical channels is reduced) at the price of higher fiber loss ($\sim$0.3~dB/km in O-band vs.\ $\sim$0.2~dB/km in C-band), which limits the reach of the QKD link. In HCF, both the lower attenuation and the drastically reduced nonlinearities (which strongly suppress Raman scattering, given the air core) enable the coexistence of quantum and classical channels in the C-band, with much longer reach.

The \emph{selective HCF placement for QKD/classical coexistence} problem~\cite{sticca2026ondm} can be stated as: \emph{given} a network topology with classical and quantum demands, an HCF budget, and a Secret Key Rate (SKR) requirement per quantum demand, \emph{decide} the HCF placement and the routing of classical and quantum channels (possibly using intermediate trusted relays of quantum modules), \emph{with the objective} of minimizing the total number of quantum modules.

\subsubsection{Methodology}
The SKR as a function of distance is computed using a standard decoy-state BB84 model (signal and decoy intensities $\mu=0.6$ and $\nu=0.1$, $N=10^{12}$ pulses, $\epsilon_\mathrm{sec}=10^{-9}$, $f_\mathrm{EC}=1.15$, 1\% misalignment, 30\% detector efficiency, $10^{-6}$ dark counts per 1-ns gate, 12.5-GHz filter) under the assumption of half/full classical loading (up to 24 co-propagating 100-GHz channels at 0~dBm/ch), and the noise contributions (ASE and SpRS, the latter 35~dB lower in HCF than in SMF~\cite{honz2023first}) are scaled with the fiber attenuation profile~\cite{sticca2026ondm}. 
We solve the network-level problem via a Genetic Algorithm that places quantum modules at the endpoints of each quantum demand and possibly at intermediate trusted relays, such that each quantum hop has a feasible SKR. Fig.~\ref{fig:example-qkd} shows the SKR versus span length for HCF ($\alpha_C=0.11$, $\alpha_O=0.2$~dB/km) and SMF ($\alpha_C=0.2$, $\alpha_O=0.3$~dB/km) under full classical loading. Taking as reach the span length at which the SKR drops below $10^{-6}$~bits/pulse, the C-band quantum channel over HCF reaches 400--480~km, against 80~km for the O-band quantum channel over SMF, i.e., the standard configuration for coexistence in SMF. This $5.6\times$ ratio is specific to the coexistence scenario: it combines the lower attenuation of the C-band HCF channel with respect to the O-band SMF channel (0.11 versus 0.3~dB/km) with the suppression of SpRS noise in the air core, which lets the quantum channel share the C-band with the classical channels. Over a dedicated dark fiber, a C-band quantum channel over SMF (0.2~dB/km) would reach well beyond 80~km, and the advantage of HCF would reduce to that of its lower attenuation.

\subsubsection{Representative results}
We consider the Tokyo metro network (23 nodes, 43 links, average link length $\sim$20km). We generate 10 quantum demands per realization, with source-destination pairs uniformly random and key-rate requests in $(2,20)$~kbps, and average over 50 independent runs. We consider two classical loadings, i.e., 50\% (\emph{half-load}) and 100\% (\emph{full-load}), the latter representing the worst case for the Raman-induced impairment of the quantum channels. 
Fig.~\ref{fig:result-qkd} shows the number of deployed quantum modules (i.e., \emph{Cost}), as a function of the HCF budget (from 0\%, i.e., \emph{SMF-only}, to 100\%, i.e., \emph{HCF-only})~\cite{sticca2026ondm}.

 
The main observations are: (a) introducing HCF reduces the number of quantum modules: since C-band coexistence over HCF extends the feasible reach of each quantum hop, fewer intermediate trusted relays are needed to serve the same demands. A 40\% HCF budget is sufficient to reduce the cost by \emph{38\%} with respect to the SMF-only baseline; (b) beyond a $\sim$40\% HCF budget, the cost saturates: the curve flattens onto the 100\%-HCF benchmark, so that upgrading further links brings no additional benefit. This confirms that a \emph{partial} HCF upgrade captures a significant share of all achievable savings; (c) the two loading conditions nearly coincide beyond $\sim$40\% HCF, whereas at low HCF budgets the full-load network is more costly than the half-load one (e.g., at 0\% HCF, the SMF-only cost is higher under full load, due to the stronger SpRS generated by the denser classical comb). Once enough links are upgraded to HCF, the quantum channels migrate to the Raman-insensitive C-band, and the dependence on the classical load diminishes.
 
The \emph{key takeaway} is that HCF is not only a capacity- or latency-oriented technology, but also an enabler for new \emph{services} (QKD and, more generally, quantum networking) that are out of reach for SMF in many practical topologies. 

\emph{Relation to other studies.} Related work on quantum communication over HCF has so far focused on the physical layer~\cite{honz2023first,liu2025ecoc_qkd}. 
To our knowledge, there are no network-level HCF placement studies for QKD/classical coexistence. 

\subsection{Scope of validity of the selective-deployment observation.} The observation that a moderate HCF budget already captures a large share of the attainable gain should not be read as a general rule. It is \emph{structural} in the edgeDC and QKD use cases, where the objective is bounded by the problem itself once every client (respectively, every quantum hop) is within reach, so that it would persist with different transponders, although the budget at which it occurs changes with the latency threshold or with the SKR requirements. It is \emph{technology-dependent} in the high-power amplification and BiDi-HCF use cases, where the benefit of HCF is realized through transponders: the budget beyond which the gain flattens, and the energy-optimal $P_{\max}$, are tied to the considered transponder sets ($\S$3.~\ref{sec:hpa} and Table~\ref{tab:txp}), and would shift with transponders supporting higher-order or probabilistically shaped formats, higher symbol rates, or a lower power per bit, with wider transmission bands, and with different traffic profiles. The recurring observation is thus that a well-chosen minority of links captures a significant share of the benefit, which makes selective deployment the natural regime while HCF remains scarce and costly, not that larger budgets are without value.

\section{Limitations, Challenges, and Opportunities}
\label{sec:challenges}

\subsection{Limitations of Current Network-Wide HCF Studies}
\label{sec:limitations}
Before discussing the broader challenges and opportunities, it is important to acknowledge the limitations of the studies presented in $\S$~\ref{sec:usecases} (and, more generally, of current network-wide HCF studies).

\textbf{Cost modeling.} The most significant limitation is the absence of a reliable HCF cost model. Because HCF is not yet a commodity, no public per-km cost data exist for production-grade deployments. Our studies (and most of the literature) either ignore fiber CAPEX or run sensitivity analyses with assumed HCF-to-SMF cost ratios. For this reason, the results of $\S$~\ref{sec:usecases} should be read in terms of the marginal benefit of additional HCF rather than as a cost-optimal budget: a given budget is economically preferable to a larger one only if the HCF cost premium exceeds the value of the additional benefit.

\textbf{Physical-layer assumptions.} The HCF transmission model~\cite{poggiolini2022jlt} assumes ideal NANF behavior in the C+L window, neglecting residual absorption peaks (e.g., from trace gases in the air core) that have been reported for current NANF prototypes~\cite{petrovich2025nature}. Similarly, our model does not capture the full impact of high-power-related phenomena (e.g., gas ionization at very high $P_{tx}$, thermal effects in the HCF coating). 

\textbf{Traffic and demand models.} We assume uniform source-destination distributions and a fixed demand mix. Real networks exhibit strong heterogeneity, with hyperscaler-dominated traffic concentrated on a subset of links between availability zones. A study with realistic traffic matrices 
could refine current deployment recommendations.


\textbf{Algorithmic scalability.} The optimization approaches adopted in these studies provide high-quality solutions across the considered network scales. Adopting such approaches for networks with thousands of links is a natural next step, for which tighter formulations, decomposition methods, and learning-based heuristics are promising directions.

\textbf{Use-case coupling.} Our four use cases are analyzed largely in isolation. In practice, a single HCF deployment must serve multiple use cases simultaneously (latency, capacity, energy, QKD), and the corresponding HCF type and placement may not be the same. A multi-objective, multi-use-case formulation is an important direction for future work.

\subsection{Open Challenges for HCF}~\label{sec:open_challenges}

\textbf{Manufacturing scale and cost.} Despite recent industrial agreements (e.g., Microsoft--Corning--Heraeus~\cite{Deployment2}), HCF production is still orders of magnitude below SMF. Reaching the economies of scale to make HCF cost-competitive with SMF on a per-Tbps basis is the key enabler for large-scale adoption.

\textbf{Standardization.} Unlike SMF, whose parameters are precisely specified by ITU-T G.652--G.657, HCF designs remain fragmented across manufacturers (NANF, DNANF, ST-HCF), with different attenuation profiles, IMI, splice losses, and mode structures. A unified framework, ideally under ITU-T and IEC, is needed for multi-vendor interoperability and for operators to write procurement specifications.
\textbf{Field deployment and operations.} HCF cables typically feature larger fiber diameters and a more complex microstructure than SMF. This implies \emph{(i)} the need for specialized splicing and connector techniques (low-loss splices down to 0.05~dB have been demonstrated with automated procedures~\cite{feng2026ofc}, but not yet at the routine, high-volume level of SMF), 
\emph{(ii)} the need for low-backscatter OTDRs for fault localization, and \emph{(iii)} the need for revised installation procedures (handling tension, bend radius, environmental sealing). Moreover, operator workforce training is a non-trivial OPEX item.

\textbf{Long-term reliability and aging.} As HCF relies on light propagation in an air-filled core, long-term reliability under realistic environmental conditions (temperature cycles, humidity, gas ingress, silica aging, micro-bend stress) must be demonstrated through multi-year lifetime tests. Some reports have shown gas-absorption features from residual water, $CO_2$, and $HCl$ introduced during fabrication, in and beyond the L-band~\cite{petrovich2025nature}. 

\textbf{High-power operation and safety.} 
The high-power regime (total EDFA output power $\sim$38~dBm) raises safety considerations
(e.g., fire risk in case of fiber breaks, eye safety for field personnel). Although these are not fundamental physical barriers, they require updated operational procedures and certifications.

\textbf{Ecosystem completeness.} HCF needs to be accompanied by a complete ecosystem of components: low-loss connectors, broadband EDFAs operating at high output power, dedicated low-backscatter OTDRs, monitoring tools (e.g., for IMI), and EMS/OSS integration. While many of these components exist as prototypes, none of them is yet a commodity.

\subsection{Opportunities for HCF Beyond the Analyzed Use Cases}
\label{sec:opportunities}

\textbf{Ultra-wideband transmission in unconventional bands.} The elimination of silica absorption peaks (water peak around 1380~nm, infrared absorption) enables HCF to transmit efficiently in unconventional bands (O, E, S), where SMF performance is severely limited. This opens the door to ultra-wideband WDM systems with effective bandwidth several times that of C+L, provided that band-specific amplifiers (e.g., TDFAs, PDFAs, BDFAs) and ROADM components are co-designed.

\textbf{High-power amplification for unrepeated systems.} As in $\S$~3\ref{sec:hpa}, HCF with high-power EDFAs enables unrepeated reaches $>$300~km~\cite{ali2025ofc}, removing inline amplifier sites, particularly attractive for greenfield and submarine deployments. 

\textbf{Single-fiber BiDi transmission.} As in $\S$~3\ref{sec:bidirectional}, the near-absence of Rayleigh backscattering doubles fiber-plant capacity at constant fiber count. 

\textbf{Quantum communication and entanglement distribution.} As in $\S$~3\ref{sec:qkd}, the air core and suppressed nonlinearity make HCF an ideal medium for QKD coexistence and, more broadly, for entanglement distribution and quantum repeaters in future quantum networks.

\textbf{Access and 6G RAN.} In Coherent-PON and 6G RAN/fronthaul scenarios, HCF offers a dual benefit: (i) the ultra-low latency directly addresses the strict eCPRI fronthaul budgets, and (ii) the high optical power budget supports passive splitting architectures with high split ratios. The combination of HCF with high-power coherent transceivers is a promising direction for next-generation access networks.

\textbf{Subsea systems.} Subsea cables are extremely sensitive to fiber CAPEX and OPEX (per-km cable cost, repeater electrical power). Ultra-low-loss HCF with high power-handling capabilities can extend repeaterless reach and reduce the number of repeaters. The mechanical and environmental requirements of subsea cables, however, are very different from terrestrial ones, and dedicated HCF cable designs are needed.

\textbf{HCF-type specialization by use case and network segment.} A peculiarity of HCF, compared to SMF, is that the optimal fiber design is not the same across use cases:
(i) latency-oriented deployments (DCI, edgeDC consolidation) prioritize \emph{propagation latency} and tolerate moderate attenuation; (ii) capacity-oriented deployments (backbone, multi-band) prioritize \emph{ultra-low attenuation} and wide bandwidth; (iii) QKD-oriented deployments prioritize \emph{ultra-low scattering and crosstalk} (e.g., low IMI, low SpRS); (iv) BiDi deployments prioritize \emph{low Rayleigh backscattering}; (v) access/PON deployments prioritize \emph{low cost} and ease of splicing.
This suggests that HCF will be deployed not as a single ``one-size-fits-all'' fiber, but as a family of fibers, each tailored to a specific network segment and use case. This is a significant departure from the SMF paradigm and has important implications for production, standardization, and procurement.

\section{Conclusion}~\label{sec:conclusion}
This paper provided an overview of \emph{emerging network-wide use cases for HCF} in next-generation optical networks. Given the current pace of industrial adoption, we argued that the relevant question for the research and operator community is no longer \emph{whether} HCF will be deployed, but \emph{where} and \emph{how} it should be deployed \emph{selectively} within existing SMF infrastructures.

Across four representative use cases: \emph{i)} edgeDC consolidation, \emph{ii)} energy-efficient high-power amplification, \emph{iii)} capacity scaling via single-fiber bidirectional transmission, and \emph{iv)} QKD/classical coexistence, a consistent qualitative message emerged: upgrading a well-chosen minority of network links (typically the most heavily traversed ``bottleneck'' links) captures a significant share of the benefit that a full HCF overhaul would deliver. 
In the near and medium term, in which HCF remains scarce and costly, selective deployment is therefore the natural deployment regime, even though it need not be a permanent optimum ($\S$4.\ref{sec:limitations}). A second cross-cutting insight is that each use case draws on a \emph{different} physical property of HCF: low latency, low nonlinearity, low Rayleigh backscattering, and Raman suppression. 


The unifying message is that HCF is likely to coexist with SMF in a heterogeneous fiber ecosystem, in which the HCF type, the placement strategy, and the optimization objectives are tightly coupled to the targeted use case and network segment. 
This calls for multi-objective, multi-use-case network-design frameworks and integrated techno-economic analysis, in close cooperation with the manufacturing, standardization, and operational communities.

\section*{Acknowledgments}

We thank Nicola Di Cicco for his contribution on the \emph{edgeDC consolidation} use case, and Alberto Gatto and Alessandro Gagliano for their contribution on the \emph{Coexistence of classical and quantum channels} use case. We thank the \emph{JOCN Editorial Board} for the invitation to the Special Issue, and the anonymous Reviewers for providing insightful comments that significantly improved the quality of this manuscript.

\bibliography{bibliography}

\begin{thebibliography}{99}

\bibitem{knight1998photonic}
J.~Knight, J.~Broeng, T.~Birks, P.~Russell,
``Photonic band gap guidance in optical fibers,''
\emph{Science}, vol.~282, pp.~1476--1478, 1998.

\bibitem{chen2024ofc}
Y.~Chen \emph{et al.},
``Hollow Core DNANF Optical Fiber with $<$0.11~dB/km Loss,''
in \emph{Proc. OFC 2024}, paper Th4A.8.

\bibitem{petrovich2025nature}
M.~Petrovich, E.~Numkam Fokoua, Y.~Chen \emph{et al.},
``Broadband optical fibre with an attenuation lower than 0.1 decibel per kilometre,''
\emph{Nature Photonics}, vol.~19, no.~11, pp.~1203--1208, 2025.

\bibitem{ding2025ecoc}
Y.~Ding \emph{et al.},
``Support Tube Hollow-Core Fiber with 0.05~dB/km Attenuation,''
in \emph{Proc. ECOC 2025}, paper Tu.04.01.

\bibitem{poggiolini2022jlt}
P.~Poggiolini and F.~Poletti,
``Opportunities and Challenges for Long-Distance Transmission in Hollow-Core Fibres,''
\emph{J. Lightw. Technol.}, vol.~40, no.~6, pp.~1605--1616, 2022.

\bibitem{Deployment2}
Microsoft Azure Networking Blog,
``Microsoft Azure Scales Hollow-Core Fiber (HCF) Production through Outsourced Manufacturing,''
September 2025.

\bibitem{microsoft_forster_2025}
C.~Forster \emph{et al.},
``The Deployment of Hollow Core Fiber (HCF) in Azure's Network,''
Microsoft Azure Networking Blog, March 2025.

\bibitem{vanvickle_ecoc25}
P.~Van Vickle (Amazon AWS),
``Invited Presentation,'' Workshop S.01.01, ECOC 2025.

\bibitem{callaghan_reinvent24}
S.~Callaghan and J.~Vasquez,
``Planet-scale networking: How AWS powers the world's largest networks,''
AWS re:Invent 2024, Session NET403, Las Vegas, NV, USA.

\bibitem{xiong2024ecoc}
Y.~Xiong \emph{et al.},
``Field-deployed hollow-core fibre cable with 0.11~dB/km loss,''
in \emph{Proc. ECOC 2024}, paper Th3B.8.

\bibitem{rizzelli_ecoc25}
G.~Rizzelli (Meta),
``Invited Presentation,'' Workshop S.02.07, ECOC 2025.

\bibitem{poggiolini2023ipc}
P.~Poggiolini, G.~Bosco, Y.~Jiang, F.~Poletti,
``The Potential for Span Length Increase with NANF,''
in \emph{Proc. IEEE IPC 2023}, pp.~1--2.

\bibitem{iqbal2021}
M.~A.~Iqbal \emph{et al.},
``First Demonstration of 400ZR DWDM Transmission through Field Deployable Hollow-Core-Fibre Cable,''
in \emph{Proc. OFC 2021}, paper F4C.2.

\bibitem{iqbal2023}
M.~A.~Iqbal, P.~Wright, A.~Lord,
``Technical Study on the Viability of Hollow-Core and Ultra-Low-Loss Silica Fibres in Metro-Core Optical Networks,''
in \emph{Proc. OFC 2023}, paper Th2A.14.

\bibitem{pedro2025}
J.~Pedro, B.~Correia, D.~Mor\~ao,
``Optimal Placement of Hollow-Core Fiber Spans to Realize Cost-Effective and High-Capacity Optical Transport Networks,''
in \emph{Proc. ECOC 2025}, paper W.02.01.106.

\bibitem{correia2025_ondm}
B.~Correia and J.~Pedro,
``Potential of Hollow-Core Fibers to Reduce Number of in-line Amplifiers in Optical Transport Networks,''
in \emph{Proc. ONDM 2025}, pp.~1--6.

\bibitem{correia2025_ofc}
B.~Correia, J.~Pedro, N.~Costa,
``Hollow-Core Fiber Specifications for Competitive Deployment in Regio/Long-Haul Optical Networks,''
in \emph{Proc. OFC 2025}, paper Tu3I.3.

\bibitem{zami2025_ofc}
T.~Zami, N.~Rossi, S.~Melle, B.~Lavigne,
``Benefit of HCF for Throughput of Future WDM Networks,''
in \emph{Proc. OFC 2025}, paper M3F.4.

\bibitem{zami2025_apc}
T.~Zami, N.~Rossi, A.~Morea, B.~Lavigne,
``How HCF vs SMF benchmarking depends on fiber and amplification parameters in transparent WDM networks,''
in \emph{Proc. Advanced Photonics Congress 2025}, paper NeTh2C.4.

\bibitem{ali2025ofc}
A.~Ali \emph{et al.},
``Unrepeated HCF Transmission over spans up to 301.7~km,''
in \emph{Proc. OFC 2025}, paper Th4A.3.

\bibitem{sticca2024ofc}
G.~S.~Sticca, M.~Ibrahimi, N.~Di~Cicco, F.~Musumeci, M.~Tornatore,
``Hollow-Core-Fiber Placement in Latency-Constrained Metro Networks with edgeDCs,''
in \emph{Proc. OFC 2024}, paper Tu3B.2.

\bibitem{sticca2024ecoc}
G.~Sticca, M.~Ibrahimi, N.~Di~Cicco, F.~Musumeci, M.~Tornatore,
``On High-Power Optical Amplification in Hollow Core Fibers for Energy Efficiency and Network Throughput Maximization,''
in \emph{Proc. ECOC 2024}, paper W2A.101, pp.~1010--1013.

\bibitem{sticca2025ofc}
G.~S.~Sticca, M.~Ibrahimi, N.~Di~Cicco, F.~Musumeci, M.~Tornatore,
``Hollow Core Fiber as a Long-Term Solution for Capacity Scaling in Optical Networks,''
in \emph{Proc. OFC 2025}, paper M3F.3.

\bibitem{sticca2026tnsm}
G.~S.~Sticca, M.~Ibrahimi, F.~Musumeci, N.~Di~Cicco, M.~Tornatore,
``Hollow-Core Fibers for Latency-Constrained and Low-Cost Edge Data Center Networks,''
\emph{IEEE Trans. Network and Service Management}, vol.~23, pp.~443--455, 2026.

\bibitem{ibrahimi2025ondm}
M.~Ibrahimi, G.~S.~Sticca, F.~Musumeci, M.~Tornatore,
``On the Benefits of Hollow-Core Fiber in Next-Generation Optical Networks [Invited],''
in \emph{Proc. ONDM 2025}, pp.~1--5.

\bibitem{sticca2026jocn}
G.~S.~Sticca, M.~Ibrahimi, P.~E.~Revelo Samaniego, F.~Musumeci, M.~Tornatore,
``Hollow-Core Fibers as an Enabler for Energy-Efficient Next-Generation Optical Networks,''
\emph{J. Opt. Commun. Netw.}, to appear, 2026.

\bibitem{sticca2026ondm}
G.~S.~Sticca, A.~Gagliano, M.~Ibrahimi, A.~Gatto, F.~Musumeci, M.~Tornatore,
``Selective Placement of Hollow-Core Fibers for QKD and Classical Communication Coexistence,''
submitted to \emph{Proc. ONDM 2026}, Munich, Germany.

\bibitem{ibrahimi2026ofcinv}
M.~Ibrahimi, G.~S.~Sticca, F.~Musumeci, M.~Tornatore,
``Emerging Network-Wide Use Cases for Hollow-Core Fibers [Invited],''
to appear in \emph{Proc. OFC 2026}.

\bibitem{suslov2021}
D.~Suslov \emph{et al.},
``Low loss, high-performance interconnection between standard single-mode fiber and antiresonant hollow-core fiber,''
\emph{Scientific Reports}, vol.~11, p.~8799, 2021.

\bibitem{semrau2019}
D.~Semrau, R.~I.~Killey, P.~Bayvel,
``A Closed-Form Approximation of the Gaussian Noise Model in the Presence of Inter-Channel Stimulated Raman Scattering,''
\emph{J. Lightw. Technol.}, vol.~37, no.~9, pp.~1924--1936, 2019.

\bibitem{poggiolini2013logon}
P.~Poggiolini \emph{et al.},
``The LOGON strategy for low-complexity control plane implementation in new-generation flexible networks,''
in \emph{Proc. OFC/NFOEC 2013}, pp.~1--3.

\bibitem{karandin2021}
O.~Karandin \emph{et al.},
``Optical Metro Network Design with Low Cost of Equipment,''
in \emph{Proc. ONDM 2021}, pp.~1--4.

\bibitem{zhang2024jocn}
Q.~Zhang, A.~Morea, P.~Layec, M.~Ibrahimi, F.~Musumeci, M.~Tornatore,
``Power-consumption analysis for different IPoWDM network architectures with ZR/ZR+ and long-haul muxponders,''
\emph{J. Opt. Commun. Netw.}, vol.~16, no.~12, pp.~1189--1203, 2024.

\bibitem{yang2025ecoc}
X.~Yang \emph{et al.},
``800G ZR Transmission Demonstration,''
in \emph{Proc. ECOC 2025}, paper Tu.01.06.2.

\bibitem{infinera_chm}
Infinera Corporation,
``Infinera CloudWave CHM2T/CHM1T Multiservice Transponder Data Sheet.''

\bibitem{sticca2023jocn}
G.~Sticca, M.~Ibrahimi, F.~Musumeci, N.~Di~Cicco, A.~Castoldi, R.~Pastorelli, M.~Tornatore,
``Selective hybrid EDFA/Raman amplifier placement to mitigate lightpath degradation in (C+L) networks,''
\emph{J. Opt. Commun. Netw.}, vol.~15, no.~8, pp.~C232--C241, 2023.

\bibitem{souza2024jocn}
A.~Souza \emph{et al.},
``Cost analysis of ultrawideband transmission in optical networks,''
\emph{J. Opt. Commun. Netw.}, vol.~16, no.~2, pp.~81--93, 2024.

\bibitem{karandin2024mf}
O.~Karandin \emph{et al.},
``Zero-cost upgrade to a multi-fiber network with partial lane-change capabilities,''
\emph{J. Opt. Commun. Netw.}, vol.~16, no.~11, pp.~H18--H26, 2024.

\bibitem{orsuti2025mcf}
D.~Orsuti \emph{et al.},
``S/C/L-Band Transmission in Few-Mode MCF With Optical Frequency Comb Regeneration,''
\emph{J. Lightw. Technol.}, vol.~43, no.~4, pp.~1786--1793, 2025.

\end{thebibliography}

\end{document}